\documentstyle[preprint,aps]{revtex}
\tightenlines
\begin{document}
\draft

\title{Electrostatic screening in fullerene molecules.}

\author{J. Gonz\'alez \\}
\address{
Instituto de Estructura de la Materia. \\
Consejo Superior de Investigaciones Cient{\'\i}ficas. \\
Serrano 123, 28006-Madrid. Spain.}

\author{F. Guinea \\}
\address{
Instituto de Ciencia de Materiales. \\
Consejo Superior de Investigaciones Cient{\'\i}ficas. \\
Cantoblanco. 28049-Madrid. Spain.}

\author{M. A. H. Vozmediano \\}
\address{
Escuela Polit\'ecnica Superior. \\
Universidad Carlos III. \\
Avda. Mediterr\'aneo 20.\\
Legan\'es. 28913 Madrid. Spain.}
\maketitle

\begin{abstract}

We study the Coulomb interactions in fullerene
molecules within a continuum formalism.
The model gives rise to
a renormalizable field theory, which
has many similarities to standard quantum
electrodynamics. The effective electric
charge at low energies is reduced by
screening processes. The associated
renormalization of the one electron Green's
function leads to the vanishing of the
quasiparticle pole. It implies the dissappearance
of coherent one particle excitations, in
close analogy to the one dimensional Luttinger
liquid. The relevance of these results
for C$_{60}$ and related molecules
is discussed.

\end{abstract}

\pacs{75.10.Jm, 75.10.Lp, 75.30.Ds.}

\narrowtext

\section{Introduction.}

The localization of the electrons within the C$_{60}$ spheres
implies that electrostatic interactions cannot be neglected
in fullerene crystals. The simplest estimate for the charging
energy of a C$_{60}$ molecule, $e^2 / R$, where $R$ is the radius,
gives a rather large value, $\approx 4$eV. Moreover,
the low density of states at the Fermi level in graphite
reduces metallic screening (see below).
In general, the intermediate size
of the molecule puts it in a regime different from
that for an isolated atom or a bulk system. Different approaches have
been used in the study of electrostatic effects
\cite{Ritchie,Dressel,Gunnarsson,Auer,Martin,Antropov,Krumm}.
It has also been proposed that they can be the
origin of superconductivity in doped systems\cite{Kivelson,TDLee}.

In the present work, we use a simple model for the
long wavelength electronic properties of fullerene
molecules\cite{PRL,Nucl}, to analyze the screening properties.
It has been shown that it gives a reasonable
approximation to the electronic levels of C$_{60}$. Furthermore,
the scheme is sufficiently general to describe
other systems with similar topology, like nanotubes and
fullerene 'onions'.
The model can also be used to study other features, like the
electron-phonon interaction with reasonable
accuracy, a shown in a previous work\cite{PRB}.

Calculations based on this model
for the electrostatic interactions between the
highest occupied orbitals (in doped systems) are in good
agreement with more detailed calculations\cite{TDLee}.

In the next section, we discuss how to incorporate
the Coulomb interaction into this model. It will
be shown to lead to a renormalizable field theory
in (2+1) dimensions. Then, we analyze in detail the leading
divergent diagrams in the theory, and
show how to deal with these divergences by
definining appropiate renormalized couplings.
The flow of these couplings, and its implication
for some quantities of physical interest,
is discussed in the next section. Finally, in
the conclusions, we discuss the applicability
of our results to C$_{60}$ and related systems,
as well as the changes induced by the finite
curvature and doping expected in these systems.

\section{The model.}

The low lying
electronic states of an isolated graphitic sheet are
well approximated by an effective Dirac equation in
(2+1) dimensions, using H\"uckel's theory of conjugated
carbon compounds. This description plays the same role as
the effective mass theories used for the study of the long
wavelength properties in other materials. The existence
of rings with an odd number of atoms in C$_{60}$ leads
to additional complications, which require the
introduction of a fictitious gauge field, to account for
the frustration that these rings induce in the electronic
states.

We use the preceding scheme to analyze first the
screening properties of an isolated graphite plane.
The single sheet, in the absence of interactions,
is semimetallic, with a vanishing density of states
at the Fermi level. Hence, the screening properties
are expected to differ significantly from those
typical of standard metals, as no screening
length, $k_{FT} = \sqrt{4 \pi e^2 N ( \epsilon_F )}$,
can be defined.

Within the approximation described above, we can write the total
hamiltonian, including the Coulomb interaction, as:

\begin{eqnarray}
 {\cal H}_{Coulomb} & = &
     \frac{3}{2} t a \int d^{2} r \; \overline{\Psi}(\mbox{\bf r})
\mbox{\boldmath $\gamma \cdot \nabla$} \Psi(\mbox{\bf r})    \nonumber  \\
 &  &  + \frac{e^{2}}{2} \int d^{2} r_{1} \int d^{2} r_{2}
\frac{\overline{\Psi}(\mbox{\bf r}_{1}) \sigma_{3}
\Psi(\mbox{\bf r}_{1})
\overline{\Psi}(\mbox{\bf r}_{2}) \sigma_{3}
\Psi(\mbox{\bf r}_{2})}{4\pi \left| \mbox{\bf r}_{1} -
 \mbox{\bf r}_{2} \right|}    \label{hcoul}
\end{eqnarray}

where the first part describes the independent electron hamiltonian
by means of the Dirac equation. The $\gamma$'s are Dirac matrices,
$t$ is the hopping between nearest neighbor orbitals,
$\sim 2.2$eV and $a$ is the interatomic distance, $\sim 1.4$\AA.
The Fermi velocity is given by $3 t a / ( 2 \hbar )$. The second part
is the standard electrostatic interaction.

We can now try to analyze the interaction term perturbatively.
To lowest order, we have the exchange term, depicted in figure (1).
Note that what we define as Coulomb interaction is not the same
as what is traditionally assumed for 2D charges, which is
a logarithmic potential. We suppose that the electronic
charge, although localized within a plane, gives rise
to a 3D electrostatic potential, which decays as $r^{-1}$.
This difference is crucial to the cutoff dependence of
the diagrams that we will encounter. For instance,
the exchange contribution to the self energy, figure (1),
can be writen as:

\begin{equation}
\Sigma_{xc} ( \vec{k} ) \propto e^2
\mbox{\boldmath $\gamma \cdot \vec{k}$}
\int_0^{\omega_c} {{d^2 k'}\over{{k'}^3}}
\end{equation}

where $\omega_c$ is an upper cutoff, above which the
linearization we use ceases to be valid. The integral in (2)
diverges logarithmically as function of its
upper and lower limits.

This divergence is an indication of the need
of some kind of renormalization procedure. The
possibility of logarithmic corrections to the
physical quantities can also be inferred from
standard power counting. If we choose units such
that velocities are dimensionless, then the
coupling constant, $e^2$, is also dimensionless.
In fact, this power counting is identical
to that in standard quantum electrodynamics.

In the following, we will account for the divergence
in equation (2), and those other divergences which
appear at higher orders in $e^2$, by renormalizing the
terms in the initial hamiltonian. A proof that it can be
done, and a detailed explanation of most technical
steps, can be found in\cite{ren}. Thus, many intermediate
steps will be omitted in this paper, athough we will
discuss in full the physical implications.

In order to simplify the calculations of the divergences,
we will couple the 2D Dirac equation in our theory
to 3D photons, which then induce the Coulomb
interaction, eq. (1). This step is not simply a
technicality, as it also allows us to discuss
retardation effects. Retardation due to the finiteness
of the velocity of light is relevant at sufficiently
large distances. This is a well known effect
in the study of Van der Waals forces, for instance.
If the typical separations, $l$, are of the
order of $\Delta \epsilon / ( \hbar c )$, where $\Delta \epsilon$
is a typical energy splitting, and $c$ the velocity
of light, then the dependence of the interaction on
$l$ changes. We study the generalization of
these interactions to graphite planes, and, as we will
see, similar effects appear.

To include the 3D electromagnetic field, we generalize the
hamiltonian, eq. (1), to:

\begin{equation}
  H  =  \frac{3}{2} t a \int d^{2} r \; \overline{\Psi}(\mbox{\bf r})
\mbox{\boldmath $\gamma \cdot \nabla$} \Psi(\mbox{\bf r})  -  e \int
d^{2} r \; j_{\mu} A^{\mu}   \label{hq}
\end{equation}

where we are using units such the velocity of light,
$c = 1$, and we are omitting the term associated to the
free electromagnetic field. It is important to realize
that this model, although formally similar to
QED, does not exhibit Lorentz invariance. We have
{\it two} dimensionless velocities, $v_F = 3 t a / ( 2 \hbar)$,
and $c$. Initially, $v_F / c \ll 1$, which corresponds
to the non relativistic limit.

The interaction of the field $A_{\mu} $ and the
electrons in the layer is described in the standard fashion,
by coupling to the conserved current:

\begin{equation}
j_{\mu } \sim \left( i \: \overline{\Psi} \gamma_0 \Psi ,
  \: i \: v \overline{\Psi} \mbox{\boldmath $\gamma $} \Psi  \right)
\end{equation}

This poses some technical problems since the electromagnetic
field propagates in three-dimensional space while we want the
dynamics of the electrons to be confined to the two-dimensional layer.
Although this may not be achieved in general, it turns out to be
possible by specializing to the Feynman gauge, which enforces
the constraint:

\begin{equation}
\nabla_{\mu} A^{\mu} = 0
\end{equation}

In order to proceed perturbatively, we need to know the
propagator of the $A_{\mu }$ field in this gauge,
which is:

\begin{equation}
\langle T A_{\mu}(t,\mbox{\bf r}) A_{\nu}(t',\mbox{\bf r}') \rangle =
-i \delta_{\mu \nu}  \int \frac{d^{4} k}{(2 \pi)^{4}}
  \frac{\mbox{\Large $e^{i {\bf k}\cdot ({\bf r} -
  {\bf r}')}$ }
\mbox{\Large $e^{-i \omega (t - t')}$ }
}{ - \omega^{2} + \mbox{\boldmath $k$}^{2} - i \epsilon }
\end{equation}

which then couples to the 2 + 1 dimensional current.
The nonrelativistic limit of
any quantity is obtained by expanding
in powers of $v_F / c$. As mentioned above, we may expect
differences between treatments based in hamiltonian (1)
and hamiltonian (3), when retardation effects become
relevant.

\section{Calculation of the leading divergences.}

We expect logarithmic divergences to all orders
in a perturbation expansion in powers of $e^2$.
To facilitate the computation, we use a dimensional
regularization procedure\cite{ren,ramond}.

The bare parameters in the hamiltonian undergo
multiplicative renormalization:

\begin{eqnarray}
v_{bare}  &  =  &  Z_{v} v_{R}   \nonumber   \\
e_{bare}  &  =  &  Z_{e} e_{R}   \nonumber   \\
\Psi_{bare}  &  =  &  Z_{\Psi}^{1/2} \Psi_{R}   \nonumber   \\
A_{bare}^{\mu}  &  =  &  Z_{A}^{1/2} A_{R}^{\mu}  \label{ren}
\end{eqnarray}

A detailed calculation of the leading divergences in each
of these factors can be found in\cite{ren}. As an
example, to lowest order in $e^2$,
we find the following contributions:

\begin{eqnarray}
Z_{\Psi} & = & 1 + \frac{e^{2}}{8 \pi^{2}} \left( 1 - 2 v^{2} \right)
\int_{0}^{1} dx \frac{\sqrt{ 1 - x }}{1 - x + v^{2}x }
\; \frac{1}{\varepsilon} \;  + \; O(e^{4})  \label{zpsi}  \nonumber \\
Z_{v} & = & 1 - \frac{e^{2}}{8 \pi^{2}}
\int_{0}^{1} dx \frac{\sqrt{ 1 - x }}{(1 - x + v^{2}x)^{2} }
\; \frac{1}{\varepsilon} \;    \nonumber      \\
  &   &  \;\;\;\;\; - \frac{e^{2}}{8 \pi^{2}} \left( 1 - 2 v^{2} \right)
\int_{0}^{1} dx \frac{\sqrt{ 1 - x }}{1 - x + v^{2}x }
\; \frac{1}{\varepsilon} \;  + \; O(e^{4})
\end{eqnarray}

where $v$ stands for the Fermi velocity in dimensions
where $c = 1$.
In the nonrelativistic limit, $v \rightarrow 0$, we can write a
an expansion in terms of $v$. The leading term in
$Z_v$ is:

\begin{eqnarray}
Z_{v} & = & 1 - \frac{1}{16 \pi^{2}} e^{2}
 \left\{ \pi \frac{1}{v} F \left(\frac{1}{2}, \frac{3}{2}; \frac{1}{2};
 v^{2} \right) - 2 \pi v \left( 1 - 2 v^{2} \right)
   F \left(\frac{3}{2}, \frac{3}{2}; \frac{3}{2};
 v^{2} \right)   \right.  \nonumber   \\
      &   &    \left. + 4 \left( 1 - 2 v^{2} \right)
F \left(1, 1; \frac{1}{2}; v^{2} \right) -
4 F \left(1, 2; \frac{3}{2};
 v^{2} \right)  \right\} \frac{1}{\varepsilon}
  +  O   \left( e^{4}   \right)    \label{alt}
\end{eqnarray}

This expression is written in terms of hypergeometric functions.

The model exhibits gauge symmetry, which is not broken
by our choice of regularization scheme. That imposes
constraints on the values of the Z's. The
renormalized Lagrangian can be written as:

\begin{eqnarray}
S_{bare}  &  =  & Z_{kin} \int dt d^{2} r \;
\overline{\Psi}( - \gamma_0
\partial_0 + Z_{v} v \mbox{\boldmath $\gamma \cdot
\nabla$}) \Psi \nonumber  \\
  &    &  - Z_{int} \; i e
\int dt d^{2} r \; \overline{\Psi}( - \gamma_0
A_0 + Z_{v} v \mbox{\boldmath $\gamma \cdot A$}) \Psi   \label{action}
\end{eqnarray}

and the gauge invariance implies that $Z_{kin}$ = $Z_{int}$.
This, in turn, implies that $Z_e = Z_A^{-1/2}$. Furthermore,
we can infer from the structure of vortex diagrams, figure (2),
that the electric charge is not renormalized. The reason
is the insertion of a $\gamma$ oerator at each vortex,
which makes the diagram tend to zero as the momentum
transferred, $q \rightarrow 0$. We have checked
that these requirements are satisfied to order $e^4$.

\section{Renormalization flow.}

Once the multiplicative constants, eq. (8) are known,
we can analyze the dependence of the propagators
upon a change in the cutoff:

\begin{equation}
\left( \Lambda \frac{\partial }{\partial \Lambda} +
 \beta_{v}(v, e^{2}) \frac{\partial }{\partial v}  +
 \beta_{e}(v, e^{2}) \frac{\partial }{\partial e}  -
         \gamma (v, e^{2})
   \right)  G(\omega , \mbox{\boldmath $k$} , \Lambda ;
   v, e^{2})  =  0
\end{equation}

where

\begin{eqnarray}
 \beta_{v} (v, e^{2})  & = &  \Lambda \frac{\partial Z_{v}
  }{\partial \Lambda } v_{R}       \label{betav}  \nonumber \\
 \beta_{e} (v, e^{2})  & = &  \Lambda \frac{\partial Z_{e}
  }{\partial \Lambda } e_{R}    \nonumber             \\
 \gamma  (v, e^{2})  & = &  \Lambda \frac{\partial
  }{\partial \Lambda }  \; log \: Z_{\Psi}
\end{eqnarray}
The well-known solution to this equation takes the form
\begin{equation}
  G(\omega, \mbox{\boldmath $k$} , \rho \Lambda ; v, e^{2} )  =
   \exp{ \left\{ \int^{\rho }  \frac{d \rho '}{\rho '} \; \gamma
\right\} }
\; \; G (\omega, \mbox{\boldmath $k$} , \Lambda ;  v_{eff}(\rho),
  e^{2}_{eff}(\rho) )           \label{rge}
\end{equation}

and the effective parameters are given by

\begin{eqnarray}
\rho \frac{\partial }{\partial \rho} v_{eff} (\rho ) & = &
 - \beta_{v} ( v_{eff},  e^{2}_{eff} )         \nonumber   \\
 \rho \frac{\partial }{\partial \rho} e_{eff} (\rho ) & = &
 - \beta_{e} ( v_{eff},  e^{2}_{eff})
\end{eqnarray}

Increasing the scale $\rho $ of the cutoff is equivalent to
measure the observables of the theory at large distance scale.
{}From the one-loop order results of the previous section we see
that $e_{eff}$ remains constant, while $v_{eff}$
does not. We expect, therefore, the effective coupling of the
electronic interaction $e^{2}/(4 \pi v)$ to have a nontrivial
renormalization group flow in the infrared regime. Actually,
from equations 7 and 12,

\begin{equation}
\rho \frac{\partial }{\partial \rho} \frac{v_{eff}}{v_{R}}
= \frac{1}{16 \pi} \frac{e^{2}}{v_{eff}}
+  O \left( \frac{e^{4}}{v^{2}_{eff}} \right) \label{flow}
\end{equation}

so that the asymptotic behaviour of the coupling is:

\begin{equation}
\frac{1}{4 \pi} \frac{e^{2}}{v_{eff}}(\Lambda / \Lambda_{0}) =
\frac{1}{4 \pi} \frac{e^{2}}{v_{0}}
 \left( 1 - \frac{1}{8 \pi}
 \frac{e^{2}}{v_{0}}
\; log \frac{\Lambda}{\Lambda_{0}} \right)^{-1/2} \label{sol}
\end{equation}

Here, $\Lambda$ is the energy at which the effective parameters
are defined.

The behavior of the coupling constants as function of
energy, eq. (\ref{sol}), is a consequence of the
screening processes which take place within the
graphite sheet. The effective coupling is reduced at
low energies and long distances. The increase in the
Fermi velocity leads to a reduction in the density
of states near the Fermi points, which can be thought
of as the tendency towards the formation of a
Coulomb gap. A true gap, however, does not exist.

The fact that the electromagnetic interaction becomes
comparatively weaker at long distances implies that
our results, although perturbative, are qualitatively
correct at low energies. The initial coupling is large
in dimensionless units (see below), but is reduced
as the renormalization proceeds. Hence, below a given scale,
our results, which have been calculated to order $e^4$,
are valid. We can not predict quantitavely,
however, the relation between the coupling at that
scale and its bare value.

The flow of the parameters towards the weak coupling
fixed point resembles the flow in QED. In fact, even a
"Landau pole" appears, if the equations are iterated
towards the ultraviolet. This pole is not too far
above the bare parameters. It probably signals
a phase transition for sufficiently large initial
couplings. We should also remark that the flow of
the Fermi velocity towards large values is bounded
by the velocity of light. At this point, the model
becomes Lorentz invariant. This symmetry should
be preserved by the subsequent renormalization flow.

The fact that the propagators acquire non
trivial exponents should be reflected in all
physical properties. For instance, the temperature
dependence of the specific heat will differ
from the $T^2$ law expected in this semimetal,
and the exponent will no longer be an integer.

The one electron propagator is of special importance,
as it measures the coherence of the quasiparticles,
and also governs the hopping processes
to other molecules or planes away from the one being
considered. The local density of states in
the non interacting case is $\hbox{Im} G ( \omega ) \sim
| \omega |$. This power law is changed to
$\hbox{Im} G ( \omega ) \sim
| \omega |^{1 + O ( e^4 / ( t a )^2 )}$. This appearance
of anomalous exponents resembles closely the features
of a one dimensional Luttinger liquid. It reflects the
same physical processes, namely, {\it the absence of
a coherent quasiparticle pole near the Fermi level}.
Thus, the system is not truly metallic, although it
does not exhibit a gap either. This result implies,
among other things, a significant reduction in the
effective hopping elements to an external system,
as discussed extensively in connection to Luttinger
liquids\cite{Wen}.

\section{Application to fullerene molecules.}

In fullerene molecules, a number of factors not considered
so far play a role:

\qquad - The initial value of the
coupling constant, $e^2/ ( t a ) \sim 4$ is large,
although in a perturbative scheme it appears divided
by numerical factors greater than unity.
As mentioned earlier, the existence of a large initial
coupling does not change qualitatively
the low energy features of the model. Screening
processes reduce this values, until a scale is
reached below which perturbation theory is
applicable. The actual exponents associated
to physical quantities, however, will depend
on the value of the coupling at this scale,
which cannot be obtained perturbatively.

\qquad - In a finite system, the typical dimensions
act like an infrared cutoff, below which
scaling is not applicable. This is a severe
restriction for the C$_{60}$ molecule, whose radius,
$R \sim 3.2$\AA is close to the distance between
carbon atoms, which is the initial cutoff. Thus,
the renormalization scheme is only a qualitative
guide to the role of electron-electron interactions in this
molecule. It suffices, however, to rule out
models based on the existence of strong metallic
screening beyond the C-C separation, like the Hubbard
model. The renormalization flow, on the other hand,
should be more precise for the larger fullerene
molecules, which may exhibit the unconventional
features reported here.

\qquad - In doped systems, the finite density
of states at the Fermi level allows for metallic
screening. The best known cases are the family
of alkali doped C$_{60}$, A$_3$C$_{60}$,where A
stands for an alkali metal. Then, the doping concentration
is 1/40 electrons per site and per spin. That
means that the Fermi is equal to
$k_F = a^{-1} \sqrt{4 \pi / ( 15 \sqrt{3}) } \sim$ 2\AA.
Using the density of states for a
linear dispersion relation, we can
define a screening length:
$k_{FT}^{-1} \sim$ 3.4 \AA. This number is similar
than the radius of the molecule, so that metallic
screening should not play a significant role.
Alternatively, we can use the density of states of K$_3$C$_{60}$
from band structure calculations\cite{Gupta}.
Its value is 25 eV$^{-1}$
per C$_{60}$ molecule. This number is higher than the estimate
given above. Hence, we find $k_{FT} \sim$ 1\AA.
However, as discussed in\cite{Gupta}, the high
density of states is due to the hybridization of a
single orbital from each C$_{60}$ molecule.
The charge transfer between C$_{60}$ molecules
in the presence of an electric field does not imply
the existence of local, intramolecular rearrangements
of the charge distribution, because the charge remains
in this triply degenerate orbital.

These points restrict the application our results, obtained
for infinite graphite sheets, to the C$_{60}$ compounds.
However, some qualitative conclusions remain valid:

\qquad - {\it The long range effects of the Coulomb
interaction are significant, and cannot be neglected.}
{}From the preceeding sections, we can conclude that
there is no screening length in isolated sheets of graphite.
The inclusion of finite size effects, and of the doping
does not change this result. The new length
scales which appear when these phenomena are taken
into account are of the order of the C$_{60}$ radius.
Thus, the effective range of the Coulomb interaction
is, at least, comparable to the size of each C$_{60}$ unit.

\qquad - {\it The absence of coherent quasiparticles
implies the existence of strong shake up effects when
electrons tunnel into the molecule.}
 The behavior of the C$_{60}$ spheres resemble
that of finite 1D Luttinger liquids. As discussed above,
the flow of the coupling constants is towards a fixed point
with no quasiparticle pole. Coherent quasiparticles cannot be
defined at this fixed point. Finite size effects halt this
flow before this situation is reached. However, its
effects will be manifested in a variety of physical processes.
The most relevant of them is, possibly, the reduction of the
effective tunneling matrix between neighboring spheres.
This is a consequence of the strong shake-up effects which are
induced when a single electron is transferred between
spheres\cite{Fisher,GU,Weiss}. The ration between the bare
intermolecular tunneling element, $t_{inter}$, and the
effective tunneling at low energies,
$t_{eff}$, goes as $t_{eff}/t_{inter} \sim ( \omega / \omega_c )^g$,
where $\omega_c$ is the initial ultraviolet cutoff, proportional
to the $\pi$ bandwidth, and $g$ is the exponent which characterizes
the anomalous scaling of the density of states at low energies.
As mentioned earlier,
this scaling low cannot be continued to energies below
a given cutoff. This cutoff is of the same order of
magnitude as $t a / R$, where $t$
is the hopping between nearest neighbor $\pi$ orbitals
on the same sphere, $a$ the interatomic distance,
and $R$ the radius of the molecule. Thus,
$\omega / \omega_c \sim a / R \sim 1 / 3$. From the fact
that we are in the strong coupling regime, $e^2 / (a t) > 1$,
we infer that $g$ is of order unity. Hence, these effects
play a significant effect in renormalizing the effective
hopping between C$_{60}$ molecules. Note that shake up effects
cannot be studied within conventional band structure calculations.

\section{Acknowledgements}

This work has been partially supported by the CICyT, Spain
(grant MAT91-0905).


\figure{Figure 1. Lowest order correction to the electron propagator
(exchange term)}

\figure{Figure 2. Lowest order correction to the electronic charge.}


\newpage
\begin{thebibliography}{99}

\bibitem{Ritchie}R. L. Hettich, R. N. Compton and R. H. Ritchie,
Phys. Rev. Lett. {\bf 67}, 1242 (1991).

\bibitem{Dressel}R. Saito, G. Dresselhaus and M. S. Dresselhaus,
Phys. Rev. B {\bf 46} 9906 (1992).

\bibitem{Gunnarsson}V. P. Antropov, O. Gunnarson and O. Jepsen,
Phys. Rev. B {\bf 46} 13647 (1992), see also
O. Gunnarson and G. Zwicknagl, Phys. Rev. Lett. {\bf 69} 957 (1992).

\bibitem{Auer}G. N. Murphy and A. Auerbach, Phys. Rev. B
{\bf 46}, 331 (1992), G. N. Murphy and A. Auerbach,
Europhys. Lett. {\bf 19} 103 (1992).

\bibitem{Martin}R. L. Martin and J. P. Ritchie,
Phys. Rev. B {\bf 48}, 4845 (1993).

\bibitem{Antropov}V. P. Antropov, O. Gunnarsson
and A. I. Liechtenstein, Phys. Rev. B {\bf 48},
7651 (1993).

\bibitem{Krumm}S. Krummacher, M. Biermann, M. Neeb,
A. Liebsch and W. Eberhardt, Phys. Rev. B {\bf 48}, 8424 (1993).

\bibitem{Kivelson}S. Chakravarty, M. Gelfand and
S. Kivelson, Science {\bf 254} 970 (1991).

\bibitem{TDLee}R. Friedberg, T. D. Lee and H. C. Ren,
Phys. Rev. B {\bf 46} 14150 (1992).

\bibitem{PRL} J. Gonz\'alez, F. Guinea and M.A.H. Vozmediano,
Phys. Rev. Lett. {\bf 69} (1992) 172.

\bibitem{Nucl} J. Gonz\'alez, F. Guinea and M. A. H.
Vozmediano, Nucl. Phys {\bf B406}, 771 (1993).

\bibitem{PRB}F. Guinea, J. Gonz\'alez and M. A. H. Vozmediano,
Phys. Rev. B, {\bf 47} 16576 (1993).

\bibitem{ren}J. Gonz\'alez, F. Guinea and M. A. H. Vozmediano,
preprint.

\bibitem{ramond} P. Ramond, {\em Field Theory. A Modern Primer.}
(Benjamin/Cummings, London, 1981).

\bibitem{Gupta} R. P. Gupta and M. Gupta,
Phys. Rev. B {\bf 47}, 11835 (1993).

\bibitem{Wen}X. G. Wen, Phys. Rev. B {\bf 42}, 6623 (1990).

\bibitem{Fisher}C. L. Kane and M. P. A. Fisher,
Phys. Rev. Lett. {\bf68} 1220 (1992).

\bibitem{GU}M. Ueda and F. Guinea, Zeits.
f\"ur Physik, {\bf 85}, 413 (1991).

\bibitem{Weiss}M. Sasetti and U. Weiss, preprint.

\end{thebibliography}
\end{document}